\documentclass{article}
\usepackage{spconf,amsmath,graphicx}
\usepackage[utf8]{inputenc}
\usepackage{chngcntr}

\usepackage{array}
\usepackage{mdwmath}
\usepackage{mdwtab}
\usepackage{eqparbox}
\usepackage{url}
\usepackage{multirow}

\usepackage{algorithm}
\usepackage{algpseudocode}
\graphicspath{{../pdf/}{../jpeg/}}
\usepackage{subcaption}
\DeclareGraphicsExtensions{.pdf,.jpeg,.png}
\usepackage{hyperref}

\title{Continuous Speech Separation with Recurrent Selective Attention Network}

\name{ \begin{tabular}{c} Yixuan Zhang$^{1,2,*}$, Zhuo Chen$^{1}$, Jian Wu$^{1}$, Takuya Yoshioka$^{1}$, Peidong Wang$^{1}$, Zhong Meng$^{1}$, Jinyu Li$^{1}$ \end{tabular}\thanks{$^{*}$This work was done during an internship at Microsoft.}}
\address{$^{1}$Microsoft, Redmond, WA, USA \\ $^{2}$The Ohio State University, Columbus, OH, USA\\
\href{mailto:zhang.7388@osu.edu}{zhang.7388@osu.edu},
\{\href{mailto:zhuc@microsoft.com}{zhuc}, \href{mailto:wujian@microsoft.com}{wujian}, 
\href{mailto:tayoshio@microsoft.com}{tayoshio}, \href{mailto:peidongwang@microsoft.com}{peidongwang}, \href{mailto:zhong.meng@microsoft.com}{zhong.meng}, \href{mailto:jinyli@microsoft.com}{jinyli}\}@microsoft.com
}

\begin{document}
\ninept
\maketitle

\begin{abstract}
While permutation invariant training (PIT) based continuous speech separation (CSS)
significantly improves the conversation transcription accuracy,  it often suffers from speech leakages and failures in separation at ``hot spot'' regions because it has a fixed number of output channels. In this paper, we propose to apply recurrent selective attention network (RSAN) to CSS, which generates a variable number of output channels based on active speaker counting. In addition, we propose a novel block-wise dependency extension of RSAN by introducing dependencies between adjacent processing blocks in the CSS framework. 
It enables the network to utilize the separation results from the previous blocks to facilitate the current block processing. 
Experimental results on the LibriCSS dataset show that the RSAN-based CSS (RSAN-CSS) network consistently improves the speech recognition accuracy over PIT-based models. The proposed block-wise dependency modeling further boosts the performance of RSAN-CSS.
\end{abstract}


\section{Introduction}
In recent years, considerable efforts \cite{watanabe2020chime, yoshioka2019advances} have been made to improve the automatic transcription of real conversations. Two key challenges in this task are the overlapped speech and quick speaker turns, which often occur during the interactions between multiple speakers. 
Speech separation, as a front-end solution for these problems, has been explored in different ways. Methods such as deep clustering (DC) \cite{hershey2016deep} and permutation invariant training (PIT) \cite{yu2017permutation, kolbaek2017multitalker}, were proposed to address the permutation issue, and they have been extended by various methods to further improve the separation quality \cite{luo2019conv, liu2019divide, luo2020dual, subakan2021attention, zeghidour2021wavesplit,yoshioka2018multi}. To process realistic long conversation recordings, which are continuous and usually partially overlapped, Yoshioka et al. \cite{yoshioka2018recognizing,yoshioka2019advances} proposed a continuous speech separation (CSS) framework. With CSS, the long recording is processed in a segment-wise manner, with a simple stitching mechanism to make the output signals consistent across the segments.  
Thus, the long-form input audio stream can be separated into several output channels with the same length, each containing continuous overlap-free utterances. 

Utterance-level PIT (uPIT) \cite{yu2017permutation}, which addresses the permutation issue by searching for the best utterance-level permutation during training, is usually used for CSS  \cite{yoshioka2018recognizing}. However, since uPIT requires a fixed number of output channels, the maximum number of separable sources is limited. 
This would be a problem for the ``hot spot'' region during a conversation, i.e., the region where the simultaneously talking speakers exceeds the pre-assumed channel number. Also, with the fixed output channel design, in the single speaker region, the speech energy sometimes leaks from the active output channel to other output channels that should have zero energies. This so-called ``leakage issue'' could lead to insertion errors during speech recognition. 

Recently, a recursive selective attention network (RSAN) \cite{kinoshita2018listening} was proposed and showed a comparable performance to uPIT for fully overlapped speech. Unlike conventional uPIT that separates each mixing source simultaneously, the recursive approach separates one source at a time and decides when to stop the recursion based on the detection of the remaining speakers in the input. Compared with uPIT, in theory, this method generates a variable number of sources, and thus it can handle an arbitrary number of overlapped speakers. In addition, this approach generates a single output channel for one active speaker region, which could help with the leakage issue.

Inspired by the promising properties of the recursive speech separation \cite{kinoshita2018listening,wang2019speech,wang2021speaker,wang2020complex,wang2021multi} and iterative speaker adaptation for speech recognition \cite{wang2018utterance,wang2018filter,wang2019bridginga,wang2019enhanced,wang2019bridging}, in our work, we examine the application of RSAN to the CSS task. In addition, we propose a novel block-wise dependency modeling in the RSAN framework for tighter integration with the CSS pipeline. In the original setup, the audio stream is separated into overlapped blocks, and each block is processed independently. Instead, we consider dependencies between the calculation in the neighboring blocks, allowing each block to reuse the previous separation results. 
The experiments on LibriCSS data set show that the RSAN-based CSS outperforms the uPIT baseline, and better performance can be achieved with the proposed block-wise dependency method.
To further explore the ``hot spot'' issue, 
we retrain RSAN with the three-speaker overlap dataset. We show that with the speaker-counting ability, the RSAN-CSS pipeline can take advantage of those more challenging overlapped training data while maintaining the performance for 1 and 2 speaker regions. In contrast, a similarly trained uPIT model suffers from significant performance degradation on the test set with fewer speech overlaps. 

The remainder of the paper is organized as follows. Sections \ref{recursive-separation} and \ref{blockwise-dependency} describe the the architecture of recurrent selective attention network and its block-wise dependency extension, respectively. In Section \ref{exp}, our experimental setup and evaluation results are described. Finally, the concluding remarks are given in Section \ref{conclusion}.

\section{Recursive Speech Separation}
\label{recursive-separation}

\subsection{Recurrent Selective Attention Network}

Our RSAN model is modified based on the original recursive selective hearing network \cite{kinoshita2018listening}. We follow the idea of performing speech separation recursively and estimating one speech source at a time. Compared with \cite{kinoshita2018listening}, instead of estimating the noise at the first iteration and speech at the following ones, we estimate speech and remaining noise at every iteration.

\begin{figure}
\centering
         \centering
         \includegraphics[width=2.7in]{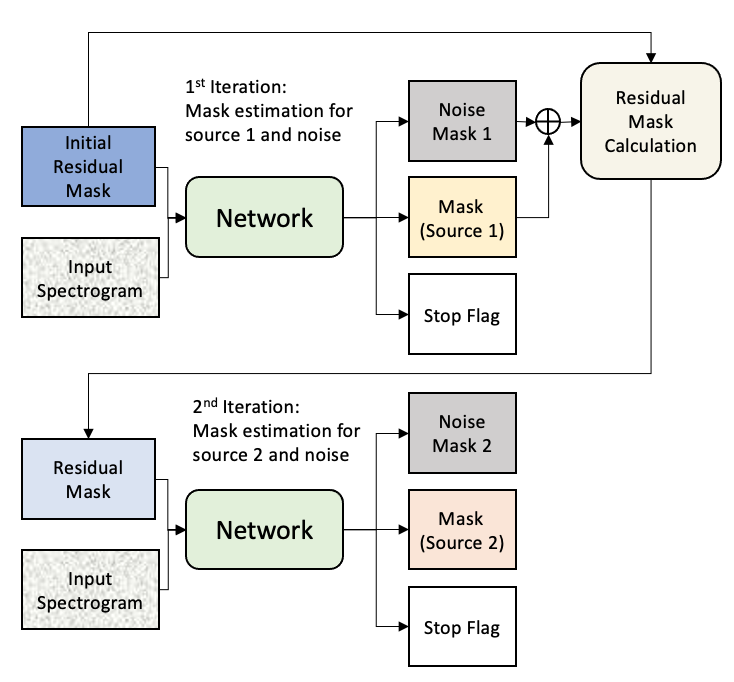}

 \caption{Illustration of speech separation with RSAN for two-speaker case.
}
\label{framework}
\end{figure}


Figure \ref{framework} explains the overall framework. The network takes the magnitude spectrogram of the input signal, and a residual mask as input.
The residual mask has the same size of the mixture spectrum, serving as an indicator of the presence of any remaining unseparated speakers in the input signal. It starts with an all-one mask, and gets updated at each iteration by subtracting the estimated masks from the current residual mask values.
In each iteration, the network generates three outputs: one mask for the extracted speaker, one mask for the remaining noise, and a stop flag.
The stop flag is a scalar within the range of  $[0,1]$. The value of the stop flag indicates the likelihood of all sources being already separated out. In other words, it indicates when the iteration should stop. During inference, a threshold is set for the stop flag. When the value of the stop flag exceeds the threshold, the iteration would stop.

An example of the two-speaker separation procedure is illustrated in Figure \ref{framework}, which consists of two iterations. 
At the first iteration, the residual mask is initialized with ones. The network thus pays attention to all regions of the input spectrogram. Before the next iteration, the residual mask is updated by subtracting the initial residual mask with the estimated masks from the first iteration. The resulting negative values in the updated residual mask, if any, are replaced with 0. When the separation is done, the noise masks from all iterations will be added together to generate the final noise mask estimate.

The training is similar to the inference procedure. One difference is that the number of iterations for each utterance is pre-determined since the total number of speakers is known for each training sample. The cost function for training is defined as,
\begin{align}
\mathcal{L} = \mathcal{L}^{\text{mse}} + \alpha \cdot \mathcal{L}^{\text{flag}},
\end{align}
where $\mathcal{L}^{\text{mse}}$ is the MSE loss between the estimated masks and the reference masks of the target sources and the noise. $\mathcal{L}^{\text{flag}}$ is the loss regarding the stop flag. $\alpha$ is set to 0.05 for all related experiments described in Section \ref{exp}. $\mathcal{L}^{\text{mse}}$ in the cost function is defined as,
\begin{equation}
\begin{aligned}
\mathcal{L}^{\text{mse}} = \frac{1}{STF} \sum_{i=1}^{S}{||\hat{\mathbf{M}_{i}} \odot \mathbf{Y} - \mathbf{A}_{\phi^*}||^2} \\
+ \frac{1}{TF} ||\hat{\mathbf{M}}_{N} \odot \mathbf{Y} - \mathbf{N}||^2,
\end{aligned}
\end{equation}
where $\mathbf{Y}, \mathbf{A}, \mathbf{N}$ are the magnitude spectrograms of the input, target sources and background noise, respectively. $\phi^*$ indicates the preferred permutation, which is determined by exhaustive search of the best permutation. $\hat{\mathbf{M}}_{i}$ is a  mask for the speaker estimated at the $i$-th iteration. $\hat{\mathbf{M}}_{N}$ is the estimated mask of the noise. $S, T, F$ correspond to the number of target sources, time frames and frequency bins. $\mathcal{L}^{\text{flag}}$ in the cost function is defined as
\begin{equation}
\mathcal{L}^{\text{flag}} = -\sum_{i=1}^{S}{[\mathbf{z}_{i}\ln{\hat{\mathbf{z}}_{i}}+(1-\mathbf{z}_{i})\ln{(1-\hat{\mathbf{z}}_{i})}]},
\end{equation}
where $\mathbf{z}$ is a vector that contains $S$ elements, in which the first $S-1$ elements are 0, and the last element is 1. $\hat{\mathbf{z}}$ is the estimated vector of the stop flags.

\subsection{Continuous Speech Separation with RSAN}

\begin{figure}
 \centering
 \includegraphics[width=3.3in]{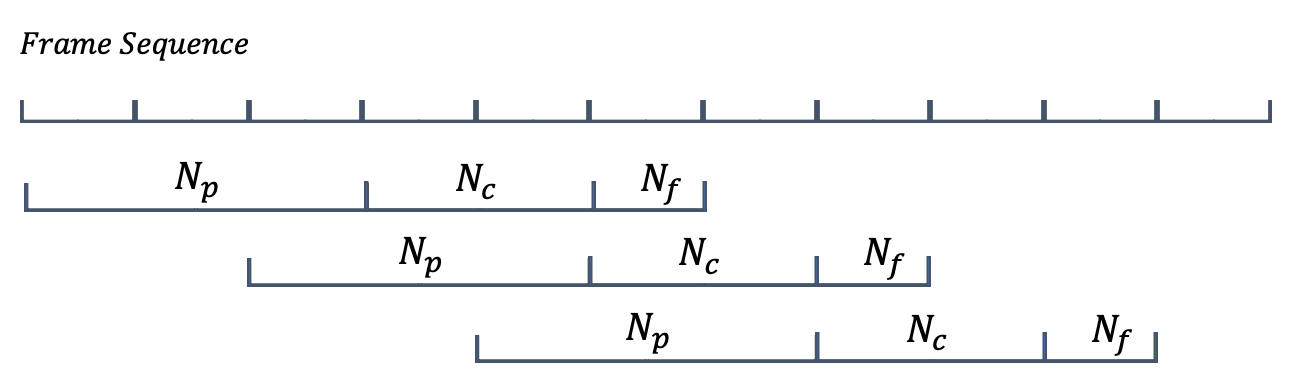}
 \caption{Streaming processing with continuous speech separation}

\label{CSS}
\end{figure}

In CSS, we first segment the long input with a sliding window and perform separation within segmented blocks.  As depicted in Figure \ref{CSS}, the input is divided into several blocks by a sliding window which consists of three parts, including $N_{p}$, $N_{c}$, $N_{f}$, each representing the past, current and future frames. Each time, the sliding window shifts by $N_{c}$ frames. 
During separation, the whole window is fed into the separation network to generate masks, and only the mask within $N_{c}$'s region is used. A stitching processing is then applied to align the permutation between adjacent windows, by comparing the mask similarity between their shared region, i.e. $N_p+N_f$. The final output is obtained by combing each aligned window.

When integrating RSAN into CSS framework, RSAN serves as a local separator for each window. As RSAN generates variable number of outputs for each block, we zero-pad the local separation result to align them with the same number of output channel.

\section{Block-wise Dependency Modeling}
\label{blockwise-dependency}

\begin{figure*}
\centering
    \begin{subfigure}[b]{0.42\linewidth}
         \centering
         \includegraphics[width=2.4in]{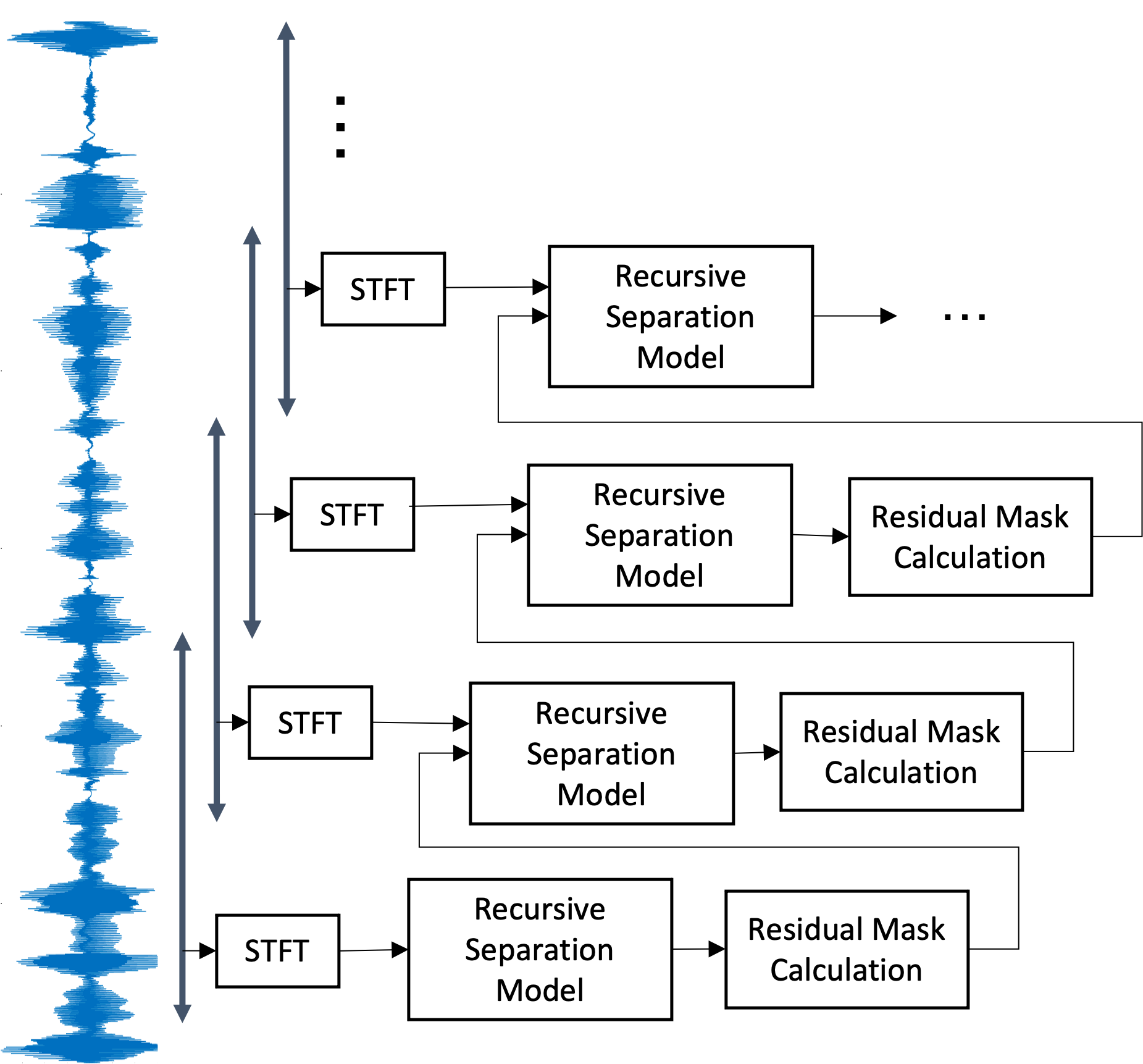}
         \caption{}
    \end{subfigure}
     \begin{subfigure}[b]{0.42\linewidth}
         \centering
         \includegraphics[width=2.9in]{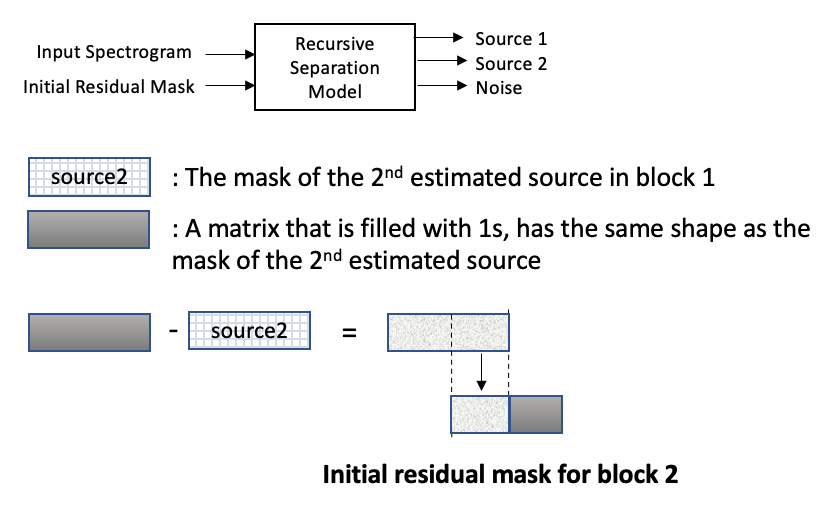}
         \caption{}
     \end{subfigure}
 \caption{Illustration of the proposed block-wise dependency modeling mechanism. (a) Overview of the dependencies between blocks. (b) An example of initial residual mask calculation. For simplicity, we plot a scenario where the previous block generates two source outputs.}
\label{blockwise}
\end{figure*}

Despite the shared content among the adjacent blocks, each block in the original CSS framework is processed independently, resulting in potentially redundant computation and sub-optimal separation due to the limited observation for each block, e.g. 1.6s in \cite{chen2020continuous}. 
In the recursive separation framework, the residual mask conveys the separation information from earlier iterations. Such message passing property can be naturally extended to sequential modeling. 
In this section, we propose a block-wise dependency mechanism to take advantage of the overlapping information between neighboring blocks by passing the separation information from the previous block to the next. This will facilitate the separation in the subsequent block.

Figure \ref{blockwise} shows the proposed block-wise dependency modeling mechanism. Figure \ref{blockwise} (a) gives an overview of how the inter-block dependencies are added to the overall CSS processing. 
The separated source masks from each block are used to calculate the initial residual mask for the next block. Regarding the first block of the long input signal, since no earlier information is available, the initial residual mask is filled with 1s.

Figure \ref{blockwise} (b) shows an example of how the initial residual mask is calculated based on the separated speech signals from the previous block. In this example, we assume that the overlap ratio between blocks is 50\%. 
To form the initial residual mask for block 2, we firstly aggregate the estimated masks from block 1. Specifically, we sum the masks from the second recursion to the end, i.e. $M_{B1}=\sum_i^{2:} M_i^{B_1}$, where $B_1$ and $B_2$ refer to block 1 and 2, $i$ indexes the recursion.  
A new mask $D$ is then constructed by subtracting $M_{B1}$ from an all one mask. 
Finally, the initial residual mask for block 2 is formed by passing the values from $D$ for the overlapped region between blocks 1 and 2, as shown below :
\begin{equation}
\begin{aligned}
     \mathbf{R}_{1}^{B2}[1:\frac{T}{2}, :] &= \mathbf{D}[\frac{T}{2}:, :] \\
    \mathbf{R}_{1}^{B2}[\frac{T}{2}:,:] &= \mathbf{1}
\end{aligned}
\end{equation}
 where $T$ is the length of the block. The newly formed residual mask $R_1^{B_2}$ contains strong bias information for the first extracted speaker in block1, thus providing a warm start for subsequent separation.
 When only 1 speaker exists in the previous block, the mask of the later estimated sources $\hat{\mathbf{M}}_{2:}^{B1}$ is filled with 0s, making $\mathbf{D}$ an all-one matrix. 

\section{Experiments}
\label{exp}

\subsection{Experimental Setup}
\label{generalsetup}
\subsubsection{Training set}
\label{trainingset}
We use the same training dataset as in \cite{chen2021continuous}, which contains 219 hours of mixed speech. 
The source utterances in each mixed sample are randomly picked from WSJ1 \cite{wsj1}.
The number of speakers in a mixture can be one or two. (We later show the experimental results with more overlapping speakers in Section \ref{subsec: more_speakers}.)
The sampled utterances are artificially reverberated by convolving with room impulse responses (RIRs) that are simulated with the image method \cite{allen1979image}.
The reverberated utterances are then rescaled and mixed to form the overlapped speech, with a random energy ratio between -5 dB to 5 dB. A simulated isotropic noise \cite{habets2007generating} clip is added to each mixture with a signal to noise ratio between 0 and 10 dB. During training, each utterance is segmented into 4s clips before feeding into the network.

\subsubsection{Evaluation details}
We evaluate the models on the LibriCSS \cite{chen2020continuous} dataset, which contains 10 hours of multi-speaker recordings. The dataset consists of 60 10-minute sessions, and
is obtained by recording the replays of the concatenated and mixed utterances from LibriSpeech \cite{panayotov2015librispeech} with a circular microphone array in the meeting room. 
We use the the first channel for evaluation. Session 0 from the LibriCSS dataset is applied as the development set for hyper-parameter tuning, e.g., the the stop flag threshold for RSAN.

The trained models are evaluated on the  conversation transcription pipeline described in \cite{yoshioka2019advances}. The asclite speaker agnostic word error rate (SA-WER) is used as the evaluation metric. During separation, the long mixture audio is segmented to 2.4s blocks with 0.8s hop size.

\subsubsection{Network training}
\label{net}

The RSAN model contains 16 Conformer encoder layers \cite{chen2021continuous}, each having 4 attention heads, 256 attention dimensions and 1024 feed forward network dimensions. The output from the last Conformer decoder layer is converted to two masks and a flag value through three corresponding sigmoid projection layers. 
The block-wise dependency model is initialized with a well trained RSAN network. Note that we use 2.4s chunk size with 0.8s hop size for block-wise dependency model training as it shows the best performance. 

For comparison, the uPIT-based model from \cite{chen2021continuous} is employed as the baseline, which has the same amount of parameters.
All models are trained using the AdamW \cite{loshchilov2017decoupled} optimizer with an initial learning rate of 0.0001. 

\subsection{Results}

\begin{table}[!t]
\caption{WER (\%) comparison of uPIT, RSAN model with or without block-wise dependency mechanism}
\label{Recursive}
\centering
\begin{tabular}{c|cccccc}
\hline
\multirow{2}{*}{Methods} & \multicolumn{6}{c}{Overlap Ratio in \%} \\ 
\cline{2-7} 
 & 0S & 0L & 10 & 20 & 30 & 40\\ \hline
uPIT  & 7.0 & 7.1 & 9.2 & 12.2 & 16.1 & 17.2 \\
RSAN  & 6.6 & 6.3 & \textbf{9.0} & 12.4 & 14.8 & 16.4 \\ 
+ Dependency & \textbf{6.3} & \textbf{6.0} & 9.1 & \textbf{11.8} & \textbf{14.1} & \textbf{15.9} \\ 
\hline
\end{tabular}
\end{table}

\begin{figure}
  \begin{center}
  \includegraphics[width=\linewidth]{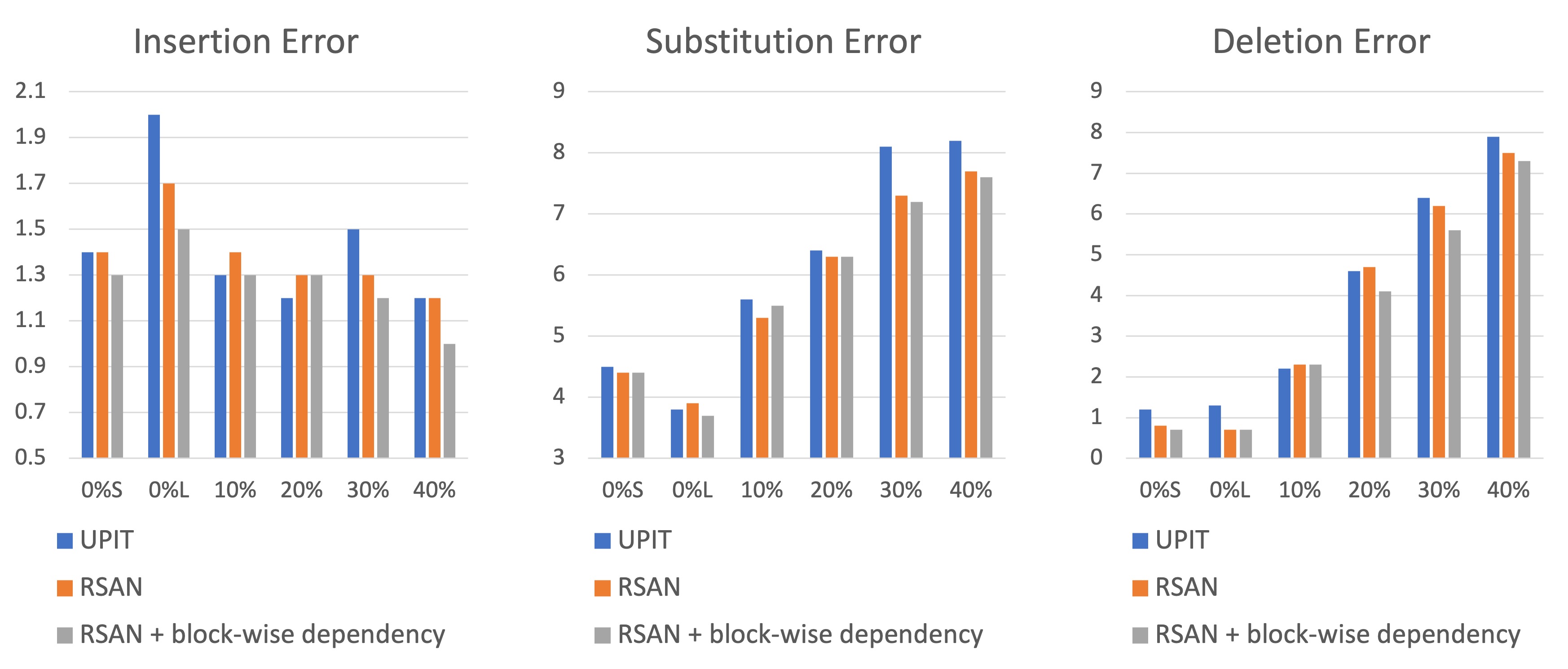}\\
  \caption{Analysis on the WER results of three methods: uPIT, RSAN and RSAN with block-wise dependency } \label{error}
  \end{center}
\end{figure}

The evaluation results are shown in Table \ref{Recursive}.
We observe that RSAN consistently outperforms uPIT. On average, in comparison with uPIT, the RSAN model reduces the WER by 5\% relatively. The improvements are mainly observed in low/high overlap scenarios such as 0L, 0S, 30\% and 40\%. Figure \ref{error} shows an analysis of the WERs of different models. Apart from the expected decrease of the insertion error due to the likely improvement on the leakage issue, it is also observed that the substitution error is reduced especially in the scenarios with high overlap ratios. A possible explanation is that the recursive separation brings better separation quality in the overlapped region, which is also observed in \cite{kinoshita2018listening}.

The average WER is further reduced with the dependency mechanism. The main improvements are from sessions that have higher overlap ratios such as 20\%, 30\% and 40\%, which shows the effectiveness of the block-wise dependency mechanism in the overlapped regions.


\subsection{CSS with more output channels}
\label{subsec: more_speakers}

In previous experiments, the evaluation is performed with 2.4s processing window, assuming that each block contains at most two active speakers. 
This assumption is not always true, especially for the ``hot spot'' regions. 
In this experiment, we relax the above assumption to having at most three speakers in a block. Accordingly, we retrain the RSAN model on a new training set which is built on the training set described in Section \ref{trainingset} by adding a third speaker (also picked from WSJ1) to some 2-speaker mixtures. Note that the retrained RSAN model doesn't have block-wise dependency. The resulting training set contains 32\% 3-speaker mixtures, 48\% 2-speaker mixtures and 20\% 1-speaker mixtures. 
A three speaker uPIT-based model is trained for comparison. The uPIT model has the same network architecture in Section \ref{net}, except that it generates 4 estimated masks, three for speakers and one for noise. 
 As LibriCSS dataset only contains overlap speech from two simultaneously talking speaker, to explore the modeling with more participants in each window, a longer window is used in the experiment.
We perform the evaluation on 2.4 and 4.8s (double-sized) audio blocks. CSS with three output channel is used in this experiment, where the empty outputing channels are zero padded for 1 speaker and 2 speaker regions.

Previously, we set one stop flag threshold for all iterations. This makes the model ``equally likely'' to stop in any iterations. Since the appearance of three or more active speakers is less likely, it is reasonable to set a lower threshold for the later iterations. In this experiment, two sets of the stop flag thresholds are used. One is setting a consistent threshold $0.6$ for all iterations. The other is setting the threshold as $0.6$ only for the first iteration, and use $0.1$ for the following iterations. In this case, the third and later iterations are not encouraged.

Experimental results are shown in Table \ref{Three}. Compared with the result in Table \ref{Recursive}, the three speaker uPIT model suffers from significant performance drop for 2.4s window. Under 4.8s window setting, the uPIT model has improved performance, but is still much worse than two speaker uPIT model. This proves our hypothesis that fixed outputting channel design negatively impacts the model generalization.

In contrast, the three speaker RSAN has the same performance for 2.4s window evaluation as its two speaker version. With longer window, a better performance is observed. A better generalization has been shown for RSAN model thanks to its speaker counting capability. 

In the evaluation of RSAN model on 2.4s audio blocks, we observe improved WERs with decreased stop flag thresholds in the later iterations. The reduction of the threshold encourages the network to stop early. In other words, the network will only estimate the third source when it's very confident, which is good since three active speakers rarely exist in small blocks. We then evaluate the model on 4.8s audio blocks, where we find that reducing the threshold for the second and later iterations increases the WERs, especially for the scenario with the highest overlap ratio. A possible explanation is that 3 speakers appear more frequently in 4.8s blocks. Decreasing the stop flag threshold for the later iterations may cause incorrect early stopping, thus results in worse WERs.

\begin{table}[!t]
\setlength{\tabcolsep}{3pt}
\caption{WER (\%) results of uPIT and RSAN models trained on three-speaker mixture dataset. $S, B$ indicate stop flag threshold and block size, respectively.}
\label{Three}
\centering
\begin{tabular}{c|c|c|cccccc}
\hline
\multirow{2}{*}{Methods} & \multirow{2}{*}{$S$} & \multirow{2}{*}{$B$} & \multicolumn{6}{c}{Overlap Ratio in \%} \\ 
\cline{4-9}
&  &  & 0S & 0L & 10 & 20 & 30 & 40 \\ 
\hline
\multirow{2}{*}{uPIT} & \multirow{2}{*}{$\times$} & 2.4s & 13.1 & 11.4 & 14.9 & 17.1 & 18.6 & 20.3 
\\ \cline{3-9}
&  & 4.8s & 10.8 & 8.0 & 12.9 & 15.4 & 17.6 & 18.7
\\ \hline
\multirow{4}{*}{RSAN} & 0.6 & 2.4s & 6.4 & 6.1 & \textbf{8.9} & 11.5 & 14.6 & 15.9 
\\ \cline{2-9}
 & [0.6, 0.1] & 2.4s & \textbf{6.3} & \textbf{5.7} & 9.0 & \textbf{11.4} & 14.7 & 15.3 
\\ \cline{2-9}
 & 0.6 & 4.8s & 6.8 & 5.8 & 9.2 & 11.8 & \textbf{14.5} & \textbf{15.2} 
\\ \cline{2-9}
 & [0.6, 0.1] & 4.8s & 7.0 & 5.8 & 9.1 & 12.0 & 14.6 & 15.9  \\ \hline
\end{tabular}
\end{table}

\section{Conclusions}
\label{conclusion}

In this paper, we explore the use of recurrent selective attention network for continuous speech separation. From the experimental results, it is observed that RSAN consistently outperforms the uPIT-based model. In addition, a novel block-wise dependency mechanism is proposed to introduce tighter connections between consecutive blocks and brings further improvement. Finally, to deal with the ``hot spot'' regions, we also retrain RSAN and uPIT model on a training set with 3-speaker mixtures.  

\clearpage
\bibliographystyle{IEEEtran}
\bibliography{Bibliography}
\end{document}